\documentclass{cta-author}

\definecolor{darkcyan}{rgb}{0.0, 0.55, 0.55}
\usepackage[colorlinks=true]{hyperref}

\usepackage[nameinlink]{cleveref}
\usepackage{extarrows}
\usepackage{multicol}
\usepackage{makecell}

\usepackage{algorithm}
\usepackage{algpseudocode}

\usepackage{xcolor}
\usepackage{pict2e}

\newsavebox{\ORCIDlogo}
\savebox{\ORCIDlogo}{%
\setlength{\unitlength}{\dimexpr 1em/256\relax}%
\begin{picture}(256,256)%
  \color[HTML]{A6CE39}\put(128,128){\circle*{256}}%
  \color{white}%
  \put(78.6,199.2){\circle*{20}}%
  \moveto(70.9,176,9)\lineto(86.3,176,9)\lineto(86.3,69.8)\lineto(70.9,69.8)%
  \closepath\fillpath%
  \moveto(108.9,176.9)\lineto(150.5,176.9)%
  \curveto(190.1,176.9)(207.5,148.6)(207.5 ,123.3)%
  \curveto(207.5,95,8)(186,69.7)(150.7,69.7)%
  \lineto(108.9,69.7)%
  \closepath\fillpath%
  \color[HTML]{A6CE39}%
  \moveto(124.3,83.6)\lineto(148.8,83.6)%
  \curveto(183.7,83.6)(191.7,110.1)(191.7,123.3)%
  \curveto(191.7,144.8)(178,163)(148,163)%
  \lineto(124.3,163)%
  \closepath\fillpath%
\end{picture}%
}

\newcommand\orcidicon[1]{\href{https://orcid.org/#1}{\usebox{\ORCIDlogo}}}

\usepackage{tikz}
\usepackage{tikzscale}
\usepackage{pgfplots}
\usepgfplotslibrary{external}
\usetikzlibrary{arrows}
\usetikzlibrary{arrows.meta}
\usetikzlibrary{positioning}
\usetikzlibrary{backgrounds}
\usepgfplotslibrary{fillbetween}
\usepackage{lmodern, tikzpeople}
\usetikzlibrary{positioning, shapes.multipart}
{}
{}
{}

\begin{document}

\supertitle{This paper is a preprint of a paper submitted to \textit{IET Blockchain} and is subject to Institution of Engineering and Technology Copyright. If accepted, the copy of record will be available at IET Digital Library}

\title{Multichain Taprootized Atomic Swaps: Introducing Untraceability through Zero-Knowledge Proofs}

\author{\au{Oleksandr Kurbatov$^{1,*}$}\orcidicon{0000-0002-8237-4377}, \au{Dmytro Zakharov$^{1}$}\orcidicon{0000-0001-9519-2444}, \au{Anton Levochko$^{1}$}\orcidicon{0009-0003-1815-7217}, \au{Kyrylo Riabov$^{1}$}\orcidicon{0009-0003-4118-8492}, Bohdan Skriabin$^{1}$\orcidicon{0009-0003-9961-5258}}

\address{\add{1}{Research Department at Distributed Lab, Kharkiv, Ukraine}
\email{ok@distributedlab.com}}

\begin{abstract}
Taprootized Atomic Swaps is an extension for Atomic Swaps that enables the untraceability of transactions in a particular swap. Based on Schnorr signatures, Taproot technology, and zero-knowledge proofs, the taprootized atomic swaps hide swap transactions between regular payments. We propose several implementation options: single-transaction protocol, multiple-transaction protocol that splits the receiving amount in an untraceable way, and multichain swap protocol. Our proposed approach works with any smart-contract-compatible chain and multiple Taproot-compatible chains. We describe the concrete implementation of the protocol and release the source code publically.
\end{abstract}

\maketitle

\section{Introduction}\label{section:introduction}

Blockchain technology has created numerous decentralized services and networks, operating according to predefined rules and using various cryptocurrencies. However, frequently, one needs to conduct certain operations \textit{between two different chains}, which is usually quite problematic. In this paper, we primarily focus on the problem of exchanging funds between several chains via an atomic and untraceable way. 

Now, imagine the setting where Alice wants to exchange $t_A$ tokens on Chain $A$ to Bob's $t_B$ tokens on Chain $B$. The most apparent option for Alice and Bob is to use centralized approaches where the mediator Carol is introduced, which: (a) takes Alice's $t_A$ tokens and sends them to Bob, (b) takes Bob's $t_B$ tokens and sends them to Alice, (c) takes some fee as a reward. However, there is an obvious reason why this scheme is entirely unsecured: Carol can steal the tokens within the swap process, so the approach works properly only if Alice and Bob trust Carol completely.

For this reason, several approaches involving mediators were developed to mitigate the issue where mediators can easily steal the tokens, such as \textit{Axelar Network} \cite{axelar}, for example. Despite the better security of such options, these methods still rely on validators. The \textit{Atomic Swaps} \cite{herlihy2018atomic} were introduced to address this issue, removing the need for third-party identities.

However, one of the core disadvantages of atomic swaps implementation in the classical form (see \cite{herlihy2018atomic,liu2018atomic,mazumdar2022towards}) is the ``digital trail'': any party can link the two transactions across blockchains where the swap occurred and find out both the participants of the swap and the proportion in which the assets were exchanged.

On the other hand, atomic swaps is a technology that initially assumed the involvement of only two parties, hence a “mathematical contract” between them directly. That is, an ideal exchange presupposes two conditions: \textit{(a)} only counterparties participate in the exchange (must-have) and \textit{(b)} only counterparties can trace the exchange (nice-to-have).

This paper describes the design of the concept of \textit{Taprootized Atomic Swaps}, with the help of which it is now possible to conceal the very fact of the swap. To an external investigator/auditor, transactions that initiate and execute atomic swaps will appear indistinguishable from regular Bitcoin payments. In the other accounting system (i.e., blockchain) involved in the transfer, more information is disclosed (the fact of the swap can be traced). However, it is impossible to link this to the corresponding Bitcoin transactions unless, obviously, the investigator has quite a specific insight from the involved parties (additional context can be provided by the time of the swap and approximate amount).

This paper focuses on implementing the protocol between EVM-compatible blockchains \cite{buterin2013ethereum} and Bitcoin \cite{nakamoto2008bitcoin} or other taproot-compatible systems. Atomic swaps offer a means to bridge the gap between these networks, enabling users to exchange Ethereum-based tokens (ERC-20 tokens \cite{somin2020erc20}) with Bitcoin and vice versa. 

However, note that this approach might be implemented for any two blockchains with the following condition: ``initiator'' chain must be taproot-compatible, while another chain should be smart-contract-compatible.

Our paper is structured as follows: first, in \Cref{section:previous-studies}, we will discuss currently existing approaches and how they differ from what we offer. In \Cref{section:prerequisites} we introduce basic cryptographic primitives, which our protocol proposals are based on. In \Cref{section:single-tx-protocol}, \Cref{section:multiple-tx-protocol}, and \Cref{section:multichain-tx-protocol}, we describe three versions of the Taprootized Atomic Swap protocol, all offering different possibilities and corresponding limitations. In \Cref{section:impl}, we outline the concrete implementation of the protocol. Finally, in \Cref{section:conclusions}, we conclude, summarizing everything described in the paper.

\section{Previous Studies}\label{section:previous-studies}

\subsection{Hashed Timelock Contract (HTLC)}

    The \textit{Hashed Timelock Contract} (HTLC), introduced in \cite{poon2016bitcoin}, implements a time-bound conditional payment. The idea is simple: the recipient must provide the secret to get designated coins in the specified timeframe; otherwise, coins can be spent by their sender.

Again, suppose Alice knows a secret value $s$ and wants to create HTLC in Bitcoin, sending $t$ BTC to Bob. To do so, Alice provides two spending paths for the transaction:
\begin{itemize}
    \item Bob shows such $x$ which satisfies $\mathsf{H}(x)=h$ where $h=\mathsf{H}(s)$ together with his signature (to prevent anyone except for Bob from spending the output). Here, $\mathsf{H}(\cdot)$ is a cryptographic hash function.
    \item After specified locktime $\ell t$, Alice can provide a signature. 
\end{itemize}
Since it is computationally infeasible to get $s$ from $h:=\mathsf{H}(s)$, there is no way Bob can spend the output if Alice has not revealed $s$. This way, if $\ell t$ time has passed, Alice can claim her tokens back. At the same time, if Bob gets $s$, he uses the first spending path and gets $t$ tokens. 

Formally, we denote such transactions by:
\begin{equation}
    \mathbf{\mathsf{T}} \gets \begin{pmatrix}
        \mathsf{versig}(\mathsf{sk}_B) \wedge \mathsf{vereq}(\mathsf{H}(x),h) \\
        \text{or} \; \mathsf{versig}(\mathsf{sk}_A) \wedge \mathsf{locktime}(\ell t)
    \end{pmatrix},
\end{equation}
where $\mathsf{versig}(\cdot)$ verifies the signature, $\mathsf{vereq}$ verifies that provided $x$ satisfies $\mathsf{H}(x)\xlongequal{} h$, and $\mathsf{lokctime}(\cdot)$ is the locktime.

Several papers have proposed an enhanced version of HTLC. For instance, \cite{tsabary2021mad} introduces the Mutual-Assured-
Destruction-HTLC (MAD-HTLC), which enhances the security by accounting for the possibility of \textit{bribery attacks}, where Alice bribes miners to delay the transaction until the timeout elapses. Additionally, \cite{sarisht2022} proposes He-HTLC that further enhances the security by accounting for active strategies and providing the Bitcoin implementation with average transaction fees. 

In this paper, we focus on the basic version described in \cite{poon2016bitcoin} since bribery attacks require significant capital and risk-tolerance, and thus are highly impractical (see \cite{Bonneau2016WhyBW} for details). However, our protocol can be easily extended to include features described in MAD/He-HTLCs.

\subsection{Atomic Swaps}

Using HTLC as the building block, atomic swaps provide a way to swap tokens between two parties without any mediator involved. Suppose that Alice, with $t_A$ tokens on chain $A$, wants to exchange her tokens with Bob, having $t_B$ tokens on chain $B$. Currently, most of the existing approaches rely on the following base algorithm, described and analyzed in detail in paper \cite{herlihy2018atomic} (and extended to multiple parties):
\begin{enumerate}
    \item Alice randomly chooses a secret $s$ and calculates $h \gets \mathsf{H}(s)$, where $\mathsf{H}(\cdot)$ is a cryptographic hash function.
    \item Alice initializes two conditions in the contract on spending $t_A$ tokens on Chain $A$: (a) pre-image of $h$ is provided, (b) locktime of $\ell t_A$ has passed. 
    \item Bob catches the transaction and retrieves $h$. Then, similarly to Alice, Bob on Chain $B$ defines two conditions of spending $t_B$ tokens: (a) pre-image of $h$ is given, (b) locktime of $\ell t_B < \ell t_A$ has passed.
    \item Alice activates the transaction on Chain $B$ and claims $t_B$ tokens. By doing so, she reveals $s$ -- the pre-image of $h$.
    \item Bob catches $h$ and claims $t_A$ tokens on chain $A$.
\end{enumerate}

\begin{figure}
\centering
\includegraphics[]{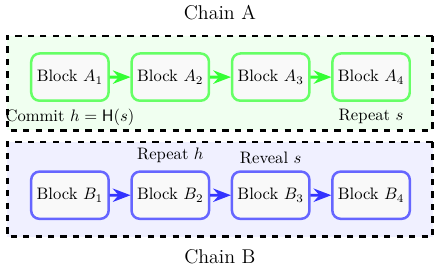}
\caption{Illustration of the classical atomic swap: (1) committing hash on chain $A$, (2) repeating the same hash on chain $B$, (3) revealing the value on chain $B$, (4) repeating the value on chain $A$.}
\label{fig:chains}
\end{figure}

As can be seen, in essence, both Alice and Bob initialize HTLC with the same hashing value $h$, but only Alice knows the pre-image of it. Finally, when Alice unlocks Bob's HTLC, she automatically enables Bob to claim tokens from her HTLC. This process is illustrated in \Cref{fig:chains}.

\section{Cryptography Prerequisites}\label{section:prerequisites}

This section will provide the basic cryptographic constructions overview needed for \textit{Taprootized Atomic Swap} protocol.

\subsection{Elliptic Curve}\label{subsection:schnorr}
Since Bitcoin natively works with \texttt{secp256k1} \cite{ecdsa}, our protocol is also based on this curve. Introduce the cyclic group $\mathbb{G}$ of prime order $q$ defined over the following elliptic curve:
\begin{equation}
    E(\mathbb{F}_p) \triangleq \{(x,y) \in \mathbb{F}_p^2: y^2=x^3 + 7\} \cup \{\mathcal{O}\},
\end{equation}
where $p=2^{256}-2^{32}-2^9-2^8-2^7-2^6-2^4-1$ is a large prime and $\mathcal{O}$ is a point on infinity, being the identity element in a group $(\mathbb{G},\boxplus)$, where $\boxplus$ is the operation of adding two points on $E(\mathbb{F}_p)$. Further assume that the group's generator is $G$, that is $qG \triangleq \underbrace{G\boxplus G \boxplus\dots\boxplus G}_{q\; \text{times}} = \mathcal{O}$. 

The security of using the group $\mathbb{G}$ on an elliptic curve is justified by the complexity of discrete log algorithm: finding the $\alpha \in \mathbb{Z}_q$ such that $\alpha G = A$ where $A \in \mathbb{G}$ is a given point on a curve. The best-known algorithm requires $O(\sqrt{q})$ group operations so that the probability attacker can find $\alpha$ given $A$ can be considered negligible (at least using classical computing).

\subsection{Schnorr Signature}

Define the hashing function $\mathsf{H}:\mathcal{M} \times \mathbb{G} \to \mathbb{Z}_q$, where $\mathcal{M}$ is the message space. The \textit{Schnorr} signature scheme $\mathcal{S}_{\text{sch}}$ is defined over functions $(G,S,V)$, where:
\begin{itemize}
    \item \textbf{Key Generation $G(1^{\lambda})$} runs by finding $\mathsf{sk} \xleftarrow[]{R} \mathbb{Z}_q,\; \mathsf{pk} \gets \mathsf{sk} \cdot G$ and outputting tuple $(\mathsf{pk},\mathsf{sk})$ -- public and private keys, respectively.
    \item \textbf{Signing Function $S(\mathsf{sk},m)$} which, based on message $m \in \mathcal{M}$ and a secret key $\mathsf{sk}=x$, conducts the following steps:
    \begin{enumerate}
        \item $r \xleftarrow[]{R} \mathbb{Z}_q$, then $X_{\sigma} \gets rG$;
        \item $c \gets \mathsf{H}(m,X_{\sigma})$;
        \item $x_{\sigma} \gets r + xc$;
        \item Output signature $\sigma := (X_{\sigma},x_{\sigma})$.
    \end{enumerate}
    \item \textbf{Verification Function $V(\mathsf{pk},m,\sigma)$}: to verify that a signature $\sigma:=(X_{\sigma},x_{\sigma})$, applied on message $m \in \mathcal{M}$, belongs to the public key $\mathsf{pk}=X$, the verification checks whether
    \begin{equation}
    x_{\sigma} G \xlongequal{?} X_{\sigma} +\mathsf{H}(m,X_{\sigma})X, 
    \end{equation}
    and if true, outputs $\mathsf{accept}$, and $\mathsf{reject}$, otherwise.
\end{itemize}

Note that sometimes, instead of using the hash function over message space and elliptic curve, one also includes the public key as the third parameter (so-called ``key-prefixed'' variant).

\subsection{zk-SNARK}\label{subsection:zksnark}

Again, fixate the finite field $\mathbb{F}_p$ of prime order $p>2$. The core considered object is a so-called \textit{circuit} -- the prover needs to show the verifier that he knows a specific secret value (called \textit{witness}), which satisfies the rule specified in this \textit{circuit}. Formally, the \textit{arithmetic circuit} is a function $C: \mathbb{F}_p^{N} \to \mathbb{F}_p^K$, being a directed acyclic graph, defining an $N$-dimensional vector of $K$-variate polynomials \cite{mayer2016}. $|C|$ is the number of gates -- the number of bilinear operations to calculate output in $\mathbb{F}_p^K$. However, usually one explicitly specify two input parameters: public statement $\mathbf{x} \in \mathbb{F}_p^n$ and witness $\mathbf{w} \in \mathbb{F}_p^m$ and the prover wants to show verifier that he knows parameters $(\mathbf{x},\mathbf{w})$ such that $C(\mathbf{x},\mathbf{w})=0$.

\begin{figure}
\centering
\includegraphics[]{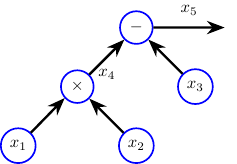}
\caption{Example of an arithmetic circuit for $C(x_1,x_2,x_3)=x_1x_2-x_3$. Both verifier and prover know the circuit, and the prover wants to show that he knows $(x_1,\dots,x_5)$ such that $x_1\times x_2=x_4$ and $x_4-x_3=x_5$}
\label{fig:circuit}
\end{figure}

A circuit example is depicted in \Cref{fig:circuit}. Here, the circuit contains $|C|=2$ gates.

To give an example of a circuit we are going to use, suppose the prover wants to show the verifier that he knows the pre-image $\mathbf{m}$ of a hash $h$ where the hash function $\mathsf{H}$ is used. In this case, the circuit is informally can be defined as $C_{\mathsf{H}}(h,\mathbf{m}) := h - \mathsf{H}(\mathbf{m})$, where all the heavy computation is encapsulated in $\mathsf{H}(\mathbf{m})$. 

Depending on which $\mathsf{H}$ is used, the different number of gates $|C|$ is used -- the smaller this number is, the better for us. For this reason, to optimize all the processes, we want to use the \textit{zk-friendly} functions, which require a smaller number of gates. As of now, the great choice is the \textit{Poseidon} hashing function \cite{Grassi2021PoseidonAN}, which natively supports messages in $\mathbb{F}_p$ and uses roughly $8\times$ fewer constraints per message bit than \textit{Pedersen} Hash \cite{pedersen}. To further clearly distinguish different hash functions, we denote the SHA256 hash function as $\mathsf{H}$, and the Poseidon zk-friendly hash as $\mathsf{H}_{\text{zk}}$.

The NARK (non-interactive argument of knowledge) is the following triple over $(S,P,V)$:
\begin{enumerate}
    \item \textbf{Setup function $S(1^{\lambda})$}: outputs public parameters $(\mathsf{pp},\mathsf{vp})$ for prover and verifier.
    \item \textbf{Proof generation $P(\mathsf{pp},\mathbf{x},\mathbf{w})$}: outputs proof $\pi$ based on public parameter $\mathsf{pp}$, public statement $\mathbf{x}$, and witness $\mathbf{w}$. 
    \item \textbf{Verifying function $V(\mathsf{vp},\mathbf{x},\pi)$}: outputs $\mathtt{accept}$ if proof $\pi$ shows that the prover knows witness $\mathbf{w}$ satisfying $C(\mathbf{x},\mathbf{w})=0$, and $\mathtt{reject}$ otherwise.
\end{enumerate}

Also, the triplet $(S,P,V)$ should satisfy the following two properties explained informally:
\begin{enumerate}
\item \textbf{Completeness}: for all $\mathbf{x} \in \mathbb{F}_p^n,\; \mathbf{w} \in \mathbb{F}_p^m$ such that $C(\mathbf{x},\mathbf{w}) = 0$:
\begin{gather}
    \mathbb{P}\left[V\left(\mathsf{vp},\mathbf{x},P(\mathsf{pp},\mathbf{x},\mathbf{w})\right)=\mathtt{accept}\right] = 1.
\end{gather}
\item \textbf{Soundness}: for any adversary prover $\mathcal{A}$, producing the proof $\pi_{\mathcal{A}}$ without knowing the witness $\mathbf{w}$,
\begin{gather}
    \mathbb{P}\left[V(\mathsf{vp},\mathbf{x},\pi_{\mathcal{A}})=\mathtt{accept}\right] = \text{negl}(\lambda)
\end{gather}
\end{enumerate}

The \textit{succinct} NARK (or SNARK for short) is the one in which the proof $|\pi|$ has the size $O_{\lambda}(\log|C|)$ and the verifying time of $O_{\lambda}(|x|,\log|C|)$. 

Finally, if we require \textbf{zero-knowledge}, the tuple $(C,\mathsf{pp},\mathsf{vp},\mathbf{x},\pi)$ reveals nothing about the witness $\mathbf{w}$.

\section{Single-transaction Protocol}\label{section:single-tx-protocol}

\subsection{Protocol flow description}
We are ready to define the concrete protocol flow. It consists of five steps, which we enumerate in the subsequent sections:
\begin{enumerate}
    \item Depositing BTC to escrow public key.
    \item Off-chain zero-knowledge proof.
    \item Depositing ETH to HTLC.
    \item Withdrawing ETH from HTLC.
    \item Spending BTC from escrow public key.
\end{enumerate}

\begin{figure*}
    \centering
    \includegraphics[]{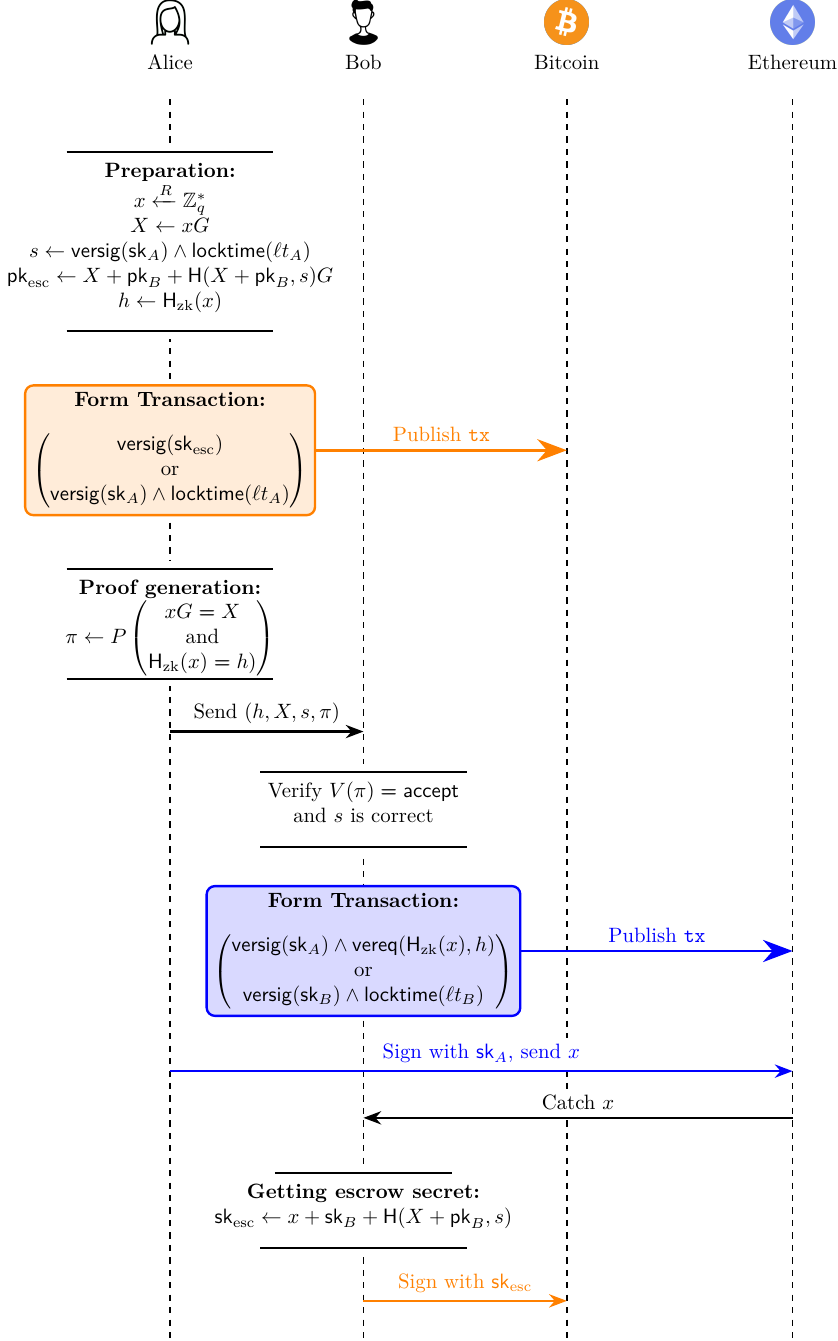}
    \caption{Single-transaction Taprootized Atomic Swap protocol}
    \label{fig:single-tx-protocol}
\end{figure*}

These steps are illustrated in \Cref{fig:single-tx-protocol}.

\subsubsection{Depositing BTC to escrow PK}

\begin{enumerate}[wide, labelwidth=1em]
    \item Alice randomly picks $x \xleftarrow[]{R} \mathbb{Z}_q^*$ and calculates public value $X \gets x \cdot G \in \mathbb{G}$.
    \item Alice forms the alternative spending path in the form of Bitcoin script $s$:
    \begin{equation}
        s \gets \mathsf{versig}(\mathsf{sk}_A)  \wedge  \mathsf{locktime}(\ell t_A),
    \end{equation}
    \item Alice calculates an escrow public key through Taproot technology:
        \begin{equation}
        \mathsf{pk}_{\text{esc}} \gets X + \mathsf{pk}_B + \mathsf{H}(X+\mathsf{pk}_B, s)\cdot G
        \end{equation}
    \item Alice calculates $h \gets \mathsf{H}_{\text{zk}}(x)$ using Poseidon hashing function.
    \item Alice forms the funding transactions and specifies the spending conditions:
    \begin{enumerate}[wide=\dimexpr\parindent+1em+\labelsep\relax, leftmargin=* ]%
        \item $\mathsf{versig}(\mathsf{sk}_{\text{esc}})$: Bob can spend the output only if he knows $x,\mathsf{sk}_B$ and script $s$.
        \item $\mathsf{versig}(\mathsf{sk}_A) \wedge \mathsf{locktime}(\ell t_A)$: Alice can spend the output, but only after a certain point in time $\ell t_A$ (this condition is the script $s$ itself).
    \end{enumerate}
     \item Alice sends the transaction to the Bitcoin network.
\end{enumerate}
\subsubsection{Off-chain zero-knowledge proof}

\begin{enumerate}
    \item Alice generates the zero-knowledge proof (see \Cref{subsection:zksnark}):
    \begin{equation}
        \pi \gets P(x\cdot G \xlongequal{} X \wedge \mathsf{H}_{\text{zk}}(x) \xlongequal{} h)
    \end{equation}
    \item Alice sends to Bob the following data:
    \begin{enumerate}
        \item $h$ -- the hash value of $x$.
        \item $X$ -- public parameter.
        \item $s$ -- alternative spending path.
        \item Generated proof $\pi$.
    \end{enumerate}
\end{enumerate}

\subsubsection{Depositing ETH to HTLC}

\begin{enumerate}
    \item Bob calculates $\mathsf{pk}_{\text{esc}}$ as:
    \begin{equation}
            \mathsf{pk}_{\text{esc},B} \gets X + \mathsf{pk}_B + \mathsf{H}(X+\mathsf{pk}_B, s) \cdot G
        \end{equation}
    and verifies the transaction that locked BTC exists (Alice might provide the transaction ID). Then Bob performs the following verifications:
    \begin{enumerate}
        \item Verifies that $V(\pi) \xlongequal{} \mathtt{accept}$, meaning Bob can access the output $\mathsf{pk}_{\text{esc}}$ if he receives $x$.
        \item Verifies that script $s$ is correct and includes only the required alternative path.
    \end{enumerate}
    \item If all the verification checks are passed, Bob forms the transaction that locks his funds on the following conditions:
    \begin{enumerate}
        \item Publishing of $x$ and the signature of $\mathsf{sk}_A$: only Alice can spend it if she reveals $x$ (the hash pre-image).
        \item $\mathsf{versig}(\mathsf{sk}_B) \wedge \mathsf{locktime}(\ell t_B)$: Bob, if he knows $\mathsf{sk}_B$, can spend the output, but only after a certain point in time $\ell t_B < \ell t_A$\footnote{Note that if $\ell t_B > \ell t_A$, Alice can firstly spend her transaction since $\ell t_A$ has passed and reveal $x$ to claim Bob's tokens. In fact, if Bob can catch the transaction in time $\Delta$, then $\ell t_B + \Delta \leq \ell t_A$.}.
    \end{enumerate}
    \item Bob sends the transaction to the Ethereum network (or any other that supports $\mathsf{H}_{\text{zk}}$).
\end{enumerate}

\subsubsection{Withdrawing ETH from HTLC}

Alice sees the locking conditions defined by Bob and publishes the $x$ together with the signature generated by her $\mathsf{sk}_A$. As a result, Alice spends funds locked by Bob. Note that if Alice does not publish the relevant $x$, Bob can return funds after $\ell t_B$ is reached.

\subsubsection{Spending BTC from escrow public key}

\begin{enumerate}
    \item If Alice publishes a transaction with $x$, Bob can recognize it and extract the $x$ value.
    \item Bob calculates the needed $\mathsf{sk}_{\text{esc}}$ as
    \begin{equation}
        \mathsf{sk}_{\text{esc},B} \gets x + \mathsf{sk}_B + \mathsf{H}(X+\mathsf{pk}_B, s)
    \end{equation}
    \item Bob sends the transaction with the signature generated by the $\mathsf{sk}_{\text{esc}}$ and spends funds locked by Alice.
\end{enumerate}

\subsection{Limitations}

Despite untraceability improvement compared to the classical Atomic Swap protocol, there are still certain drawbacks, some of which we address in the following sections:

\begin{itemize}
    \item \textbf{Matching amounts in blockchains within some time range:} the external auditor can see how many coins/tokens were withdrawn from the appropriate contract and try to find the transaction in the Bitcoin networks that pay BTC (or assets issued in the Bitcoin system) in the corresponding ratio (based on market prices). If the payment was not instant -- the auditor can assume the time range in which the swap was performed (the lock-time in the contract can provide more info about it) and select all suitable transactions. Potentially, this number can be large, but it still simplifies building the graph of transactions for auditors with specialized equipment. 
    \item \textbf{zk-friendly hashing function support:} as we have seen, the ``Bob''s chain should support the zk-friendly hashing function $\mathsf{H}_{\text{zk}}$. This is not always possible, and although non-zk-friendly functions can still be used (like SHA1 or SHA256), the corresponding circuits are much less efficient.
    \item \textbf{Secret proper management:} the secret $x$ should be managed appropriately and caught on time by Bob to avoid losing money.
\end{itemize}

\section{Multiple-transaction Protocol}\label{section:multiple-tx-protocol}

\subsection{Motivation}
This section will provide a concept of breaking the amount of BTC that must be transferred into several transactions that will be processed via atomic way (not simultaneously, but within the timelock interval). It increases the difficulty of matching swap transactions because, in this case, the external auditor needs to solve a sudoku puzzle with a much larger number of combinations than direct swap transactions.

This solution can be applied to the swap between a smart contract platform and several chains that support Taproot technology. This way, the untraceability of swaps will be significantly increased, but the user experience will be more complicated.

\subsection{Protocol Flow}

\begin{figure*}
    \centering
    \includegraphics[]{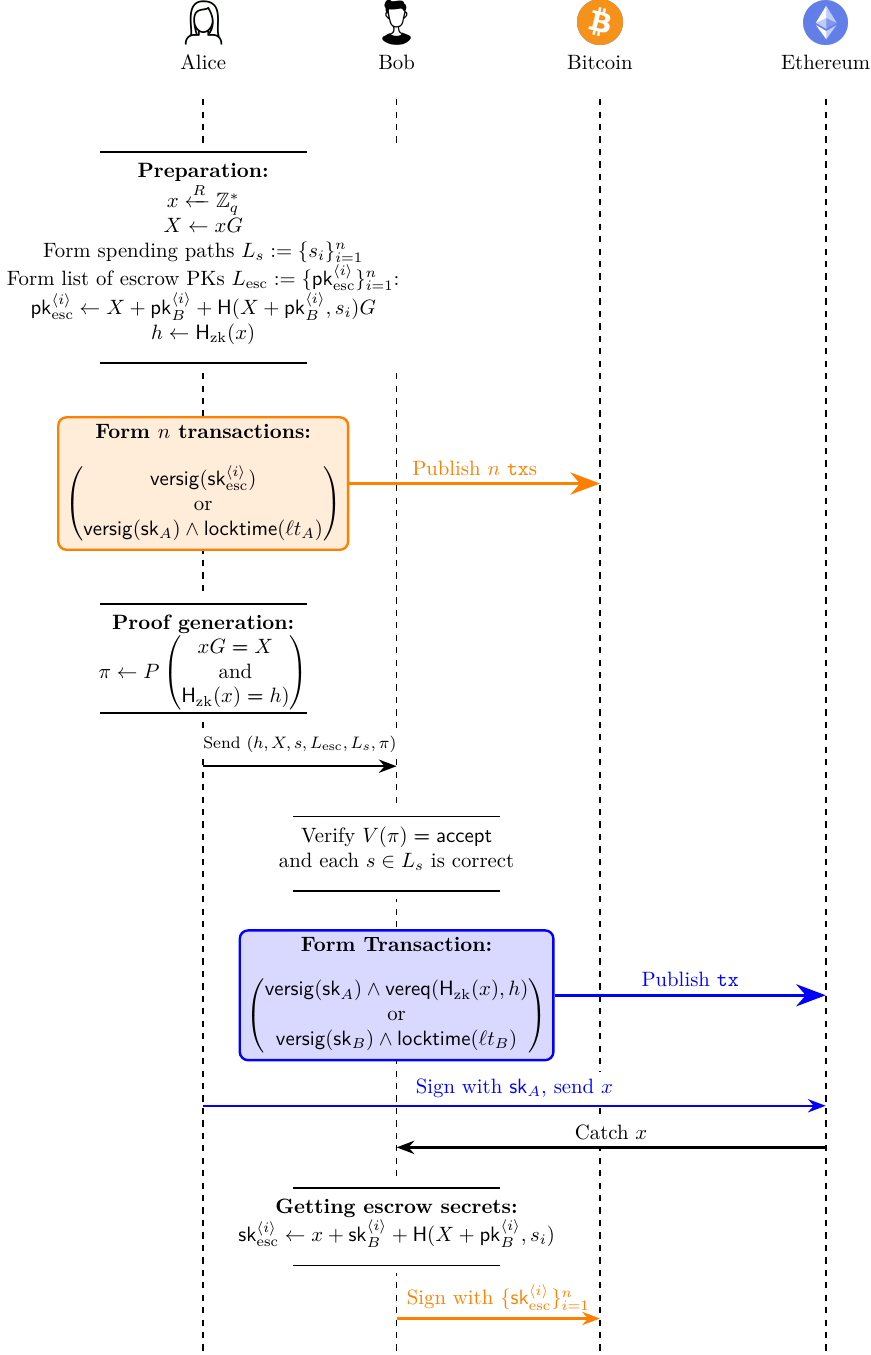}
    \caption{Multiple-transaction Taprootized Atomic Swap protocol}
    \label{fig:multiple-tx-protocol}
\end{figure*}

The flow is illustrated in \Cref{fig:multiple-tx-protocol}. The extension is the following — instead of forming a single $\mathsf{pk}_B$ by Bob, he can generate the array $\{\mathsf{pk}_B^{\langle i \rangle}\}_{i=1}^n$ according to the number $n$ of transactions Alice wants to spend. Also, further we use notation $[n]$ to denote the set $\{1,\dots,n\}$. 

The updated protocol works the following way:
\begin{enumerate}
    \item Alice has $t$ BTC on separate outputs $\{t_i\}_{i=1}^n$ (that is, $\sum_{i=1}^n t_i=t$).
    \item Alice picks $x \xleftarrow[]{R} \mathbb{Z}_q^*$ and calculates $X \gets xG$. She also forms the alternative spending paths $L_s:=\{s_i\}_{i=1}^n$. 
    \item Alice calculates the array of escrow public keys $L_{\text{esc}} := \{\mathsf{pk}_{\text{esc}}^{\langle i \rangle}\}_{i=1}^n$ as follows:
    \begin{equation}
        \mathsf{pk}_{\text{esc}}^{\langle i \rangle} \gets X + \mathsf{pk}_B^{\langle i \rangle} + \mathsf{H}(X+\mathsf{pk}_B^{\langle i \rangle},s_i)\cdot G, \; i \in [n]
    \end{equation}
    and the hash value $h \gets \mathsf{H}_{\text{zk}}(x)$. 
    \item Alice sends the list of transactions $\{\mathbf{\mathsf{T}}_i\}_{i=1}^n$ with the spending conditions:
    \begin{equation}
        \mathbf{\mathsf{T}}_i \gets \begin{pmatrix}\mathsf{versig}(\mathsf{sk}_A) \wedge \mathsf{locktime}(\ell t_A) \\ \text{or } \mathsf{versig}(\mathsf{sk}_{\text{esc}}^{\langle i \rangle})
        \end{pmatrix}, \; i \in [n]
    \end{equation}
    \item Alice generates the proof:
    \begin{equation}
        \pi \gets P\left(\mathsf{H}_{\text{zk}}(x)\xlongequal{} h \wedge xG \xlongequal{} X\right),
    \end{equation}
    and sends values $X,L_{\text{esc}},L_s,h,\pi$ to Bob (recall that $L_{\text{esc}}$ is a list of formed escrow public keys while $L_s$ is a list of alternative spending paths).
    \item Bob verifies that (a) each script $s \in L_s$ is correct and includes only the required corresponding alternative path, and (b) $V(\pi) \xlongequal{} \mathsf{accept}$. Then, he locks his tokens to the smart contract with the following conditions:
    \begin{enumerate}
        \item Publishing of $x$ and checking $\textsf{vereq}(h,\mathsf{H}_{\text{zk}}(x))$.
        \item $\mathsf{versig}(\mathsf{pk}_B) \wedge \mathsf{locktime}(\ell t_B)$. 
    \end{enumerate}
    \item Alice withdraws ETH by providing $x$.
    \item Bob takes $x$ and generates secret escrow keys as follows:
    \begin{equation}
        \mathsf{sk}_{\text{esc}}^{\langle i \rangle} \gets x + \mathsf{sk}_B^{\langle i \rangle} + \mathsf{H}(X+\mathsf{pk}_B^{\langle i \rangle},s_i)\cdot G, \; i \in [n],
    \end{equation}
    which he uses to claim amounts $\{t_i\}_{i=1}^n$.
\end{enumerate}

\section{Multichain TAS protocol}\label{section:multichain-tx-protocol}

\subsection{Motivation}
Besides the better anonymity proposed in \Cref{section:multiple-tx-protocol}, we can do much more. We can easily extend the protocol to multiple networks supporting Taproot technology! Imagine that Alice and Bob agreed to change 20 ETH to 0.8 BTC and 3 LTC. With Taptootized Atomic Swaps, they can do that via atomic way.

\subsection{Protocol Flow}

\begin{figure*}
    \centering
    \includegraphics[scale=0.9]{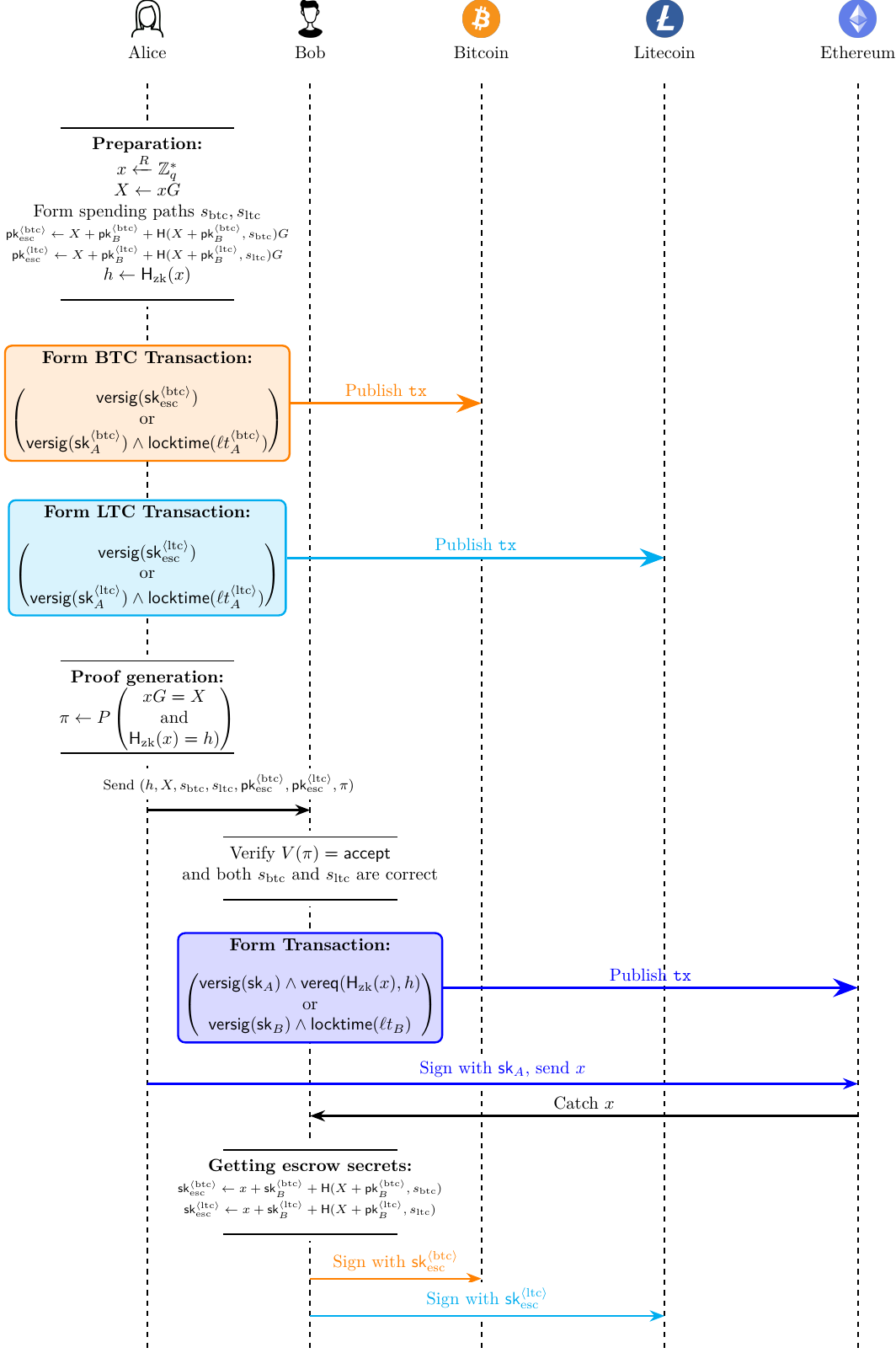}
    \caption{Multichain Taprootized Atomic Swap protocol}
    \label{fig:multi-chain-protocol}
\end{figure*}

The flow is illustrated in \Cref{fig:multi-chain-protocol}. We conduct the following steps:

\begin{enumerate}
    \item Alice with BTC keypair $(\mathsf{pk}_{A}^{\langle \text{btc} \rangle},\mathsf{sk}_{A}^{\langle \text{btc} \rangle})$ and LTC keypair $(\mathsf{pk}_{A}^{\langle \text{ltc} \rangle},\mathsf{sk}_{A}^{\langle \text{ltc} \rangle})$ has $t_{\text{btc}}$ BTC and $t_{\text{ltc}}$ LTC. Bob needs to generate two keypairs ($n=2$), the first one for the payment on Bitcoin network $(\mathsf{pk}_{B}^{\langle \text{btc} \rangle},\mathsf{sk}_{B}^{\langle \text{btc} \rangle})$ and the second one for the payment on Litecoin $(\mathsf{pk}_{B}^{\langle \text{ltc} \rangle},\mathsf{sk}_{B}^{\langle \text{ltc} \rangle})$. He wants to exchange Alice's funds for $t_{\text{eth}}$ ETH tokens.
    \item Alice picks $x \xleftarrow[]{R}\mathbb{Z}_q^*$ and calculates $X \gets xG$. She also forms the alternative spending paths $s_{\text{btc}}$ and $s_{\text{ltc}}$ for Bitcoin and Litecoin transactions, respectively.
    \item Alice calculates two values:
    \begin{gather}
        \mathsf{pk}_{\text{esc}}^{\langle \text{btc} \rangle} \gets X + \mathsf{pk}_{B}^{\langle \text{btc}\rangle} + \mathsf{H}(X + \mathsf{pk}_{B}^{\langle \text{btc}\rangle},s_{\text{btc}})\cdot G, \nonumber\\ \mathsf{pk}_{\text{esc}}^{\langle \text{ltc} \rangle} \gets X + \mathsf{pk}_{B}^{\langle \text{ltc}\rangle} + \mathsf{H}(X + \mathsf{pk}_{B}^{\langle \text{ltc}\rangle},s_{\text{ltc}})\cdot G.
    \end{gather}
    \item Alice broadcasts transactions to Bitcoin and Litecoin networks (locktimes might be different):
    \begin{enumerate}
        \item $\mathbf{\mathsf{T}}_{\text{btc}} \gets \begin{pmatrix}
            \mathsf{versig}(\mathsf{sk}_A^{\langle \text{btc} \rangle}) \wedge \mathsf{locktime}(\ell t_A^{\langle \text{btc} \rangle}) \\ \text{or}\;\mathsf{versig}(\mathsf{sk}_{\text{esc}}^{\langle \text{btc} \rangle})
        \end{pmatrix}$
        \item $\mathbf{\mathsf{T}}_{\text{ltc}} \gets \begin{pmatrix}
            \mathsf{versig}(\mathsf{sk}_A^{\langle \text{ltc} \rangle}) \wedge \mathsf{locktime}(\ell t_A^{\langle \text{ltc} \rangle}) \\ \text{or}\;\mathsf{versig}(\mathsf{sk}_{\text{esc}}^{\langle \text{ltc} \rangle})
        \end{pmatrix}$
        \end{enumerate}
    \item Alice generates the proof:
    \begin{equation}
        \pi \gets P(\mathsf{H}_{\text{zk}}(x) = h \wedge xG=X).
    \end{equation}
    \item Alice sends $(X,\mathsf{pk}_{\text{esc}}^{\langle\text{btc}\rangle},\mathsf{pk}_{\text{esc}}^{\langle\text{ltc}\rangle},s_{\text{btc}},s_{\text{ltc}},h,\pi)$.
    \item Bob asserts $V(\pi) \xlongequal{} \mathsf{accept}$, verifies that scripts $s_{\text{btc}}$ and $s_{\text{ltc}}$ are correct, and then locks his $t_{\text{eth}}$ with the spending conditions:
    \begin{enumerate}
        \item publishing of $x$ and checking $\mathsf{vereq}(h,\mathsf{H}_{\text{zk}}(x))$;
        \item $\mathsf{versig}(\mathsf{pk}_B) \wedge \mathsf{locktime}(\ell t_{\text{eth}})$ (note that $\ell t_{\text{eth}}$ must be smaller than both $\ell t_{\text{btc}}$ and $\ell t_{\text{ltc}}$).
    \end{enumerate}
    \item Alice reveals $x$ and claims her $t_{\text{eth}}$ tokens.
    \item Bob catches $x$ and calculates:
    \begin{gather}
        \mathsf{sk}_{\text{esc}}^{\langle \text{btc} \rangle} \gets x + \mathsf{sk}_B^{\langle \text{btc} \rangle} + \mathsf{H}(X + \mathsf{pk}_B^{\langle \text{btc} \rangle},s_{\text{btc}}), \nonumber\\ \mathsf{sk}_{\text{esc}}^{\langle \text{ltc} \rangle} \gets x + \mathsf{sk}_B^{\langle \text{ltc} \rangle} + \mathsf{H}(X + \mathsf{pk}_B^{\langle \text{ltc} \rangle},s_{\text{ltc}})
    \end{gather}
    to claim his $t_{\text{btc}}$ and $t_{\text{ltc}}$ tokens.
\end{enumerate}

\section{Implementation}\label{section:impl}

\begin{table*}
    \centering
    \caption{Mainnet deployment parameters}
    \begin{tabular}{cc}
         \Xhline{3\arrayrulewidth}
        \textbf{Parameter} & \textbf{Value} \\
        Alice locks BTC $\mathsf{tx}$ & $\mathsf{850e9258bf8b3bb280d32a647198d8024aece543dc283f7bfa526f4c0ceb1ab8}$ \\
        Bob locks ETH $\mathsf{tx}$ & $\mathsf{723919c0e8ec57d38792ec29b2cb82ee885b9fbbc886b34ff40fb5d3f7cc9b43}$ \\
        Alice withdraws ETH $\mathsf{tx}$ & $\mathsf{47546191a7c99ec4a7ddc243d6ea75d345ab3aff0762e09dd2f537731bd484f3}$ \\
        Bob spends BTC $\mathsf{tx}$ & $\mathsf{859dbfaa901d7106aecc8cb29966ede0c9d7a17c2cae31f4d420c1d770e9706d}$ \\
        ETH mainnet contract address & $\mathsf{0x936f971455bc674F77312f451963681fe964E838}$ \\
        \Xhline{3\arrayrulewidth}
    \end{tabular}
    \label{tab:deploy}
\end{table*}

The code\footnote{\href{https://github.com/distributed-lab/taprootized-atomic-swaps}{https://github.com/distributed-lab/taprootized-atomic-swaps}} is mainly written in Rust, because of its efficiency and easy-to-use suite of libraries, such as BDK, that provides everything to build a Bitcoin wallet and create/spend UTXOs with custom spending conditions, e.g., Taproot. For the witness generation, there are two options (code of both can be easily obtained by \textit{Iden3 SnarkJS command line utility}): either use \textit{WASM} code and execute it in its runtime or use C++ bindings. The first, and chosen, option is a way of easy implementation and flexibility because there is no need to recompile the entire application to change the witness generator code. Still, it is significantly slower than its competitor (11 times in this case). To generate the zero-knowledge proof, the bindings to the \textit{Iden3 Rapidsnark} C++ library\footnote{\href{(https://github.com/iden3/rapidsnark}{(https://github.com/iden3/rapidsnark}} have been used by the reason of its proven reliability and efficiency. The \textit{Arkworks Groth16}, a Rust-based native implementation of the \textit{Groth16 zk-SNARK}, is utilized for verifying zero-knowledge proofs.

The zk-SNARK circuits are developed using \textit{Circom} \cite{circom}, and \textit{0xPARC's circom-ecdsa}\footnote{\href{https://github.com/0xPARC/circom-ecdsa}{https://github.com/0xPARC/circom-ecdsa}} implementations, while the EVM-compatible contracts are crafted in \textit{Solidity}.

We also decided to test the performance of the zero-knowledge proof: generation time, verification time, and size of a proving key. After trying on \textit{M1 Pro Macbook}, we got parameters specified in \Cref{tab:zkparams}.

We also tested the Single-transaction protocol on Bitcoin and Ethereum mainnets! The transactions are specified in \Cref{tab:deploy}.

\begin{table}
    \centering
    \caption{Zero-knowledge proof setup parameters}
    \begin{tabular}{cc}
         \Xhline{3\arrayrulewidth}
        \textbf{Parameter} & \textbf{Value} \\
        Witness generation time, WASM & $\approx 11\text{s}$ \\
        Witness generation time, C++ & $\approx 1\text{s}$ \\
        Proof generation time & $1.51\text{s}$ \\
        Proof verification time & $\ll 1\text{s}$ \\
        Proving key size & 107MB \\
        Non-linear constraints \# & $\approx 95.7\text{k}$ \\
        Public outputs \# & $9$ \\
        Private inputs \# & $4$ \\
        \Xhline{3\arrayrulewidth}
    \end{tabular}
    \label{tab:zkparams}
\end{table}

\section{Conclusions}\label{section:conclusions}

In this paper, we presented the novel multichain anonymous atomic swap protocol, which conceals the very fact of an exchange while preserving properties of the classical Atomic Swap \cite{herlihy2018atomic}. We proposed three different versions of the protocol: the basic standard version with two spending transactions involved, the multiple-transaction version where we additionally conceal the ratio of swapped funds, and finally, the multichain taprootized atomic swap with an ultimate goal of developing technology for multi-chain swaps (i.e., smart contracts that operate on multiple chains).

In summary, we get the following advantages of the proposed multichain Taprootized Atomic Swap protocol:
\begin{enumerate}
    \item Auditors cannot match swap transactions based on \textbf{committed hash values} and appropriate \textbf{secrets} like in classic atomic swaps.
    \item Auditors cannot assume if the particular Bitcoin transaction participates in the swap — it is masked as a \textbf{regular payment transaction}. The locktime condition is hidden in the Taproot and revealed only if the swap was not performed.
    \item Auditors \textbf{cannot match amounts} in chains directly if the split mechanism is used. However, sudoku analysis can be applied to make some assumptions.
    \item The protocol is \textbf{trustless}. The protocol guarantees that only the publishing of secret k can unlock money from the contract. At the same time, publishing $x$ leads to the ability to form the correct key and produce the signature for BTC unlock.
    \item \textbf{No mediator} is required. Users can exchange the needed information for the swap directly, using existing protocols for secure message transfer.
    \item The protocol works \textbf{not only for the native currencies} but also supports tokenized assets, non-fungible tokens, etc. It can be a basic protocol for bridges, stablecoins, marketplaces, etc.
\end{enumerate}

Finally, we provided the practical implementation with a detailed rationale behind the chosen technology stack and conducted the first Taprootized Atomic Swap on Bitcoin and Ethereum mainnets. Despite some remaining challenges, like better secrets management, a user-friendly frontend, and the cost of multiple transactions, our protocol is ready for real-world use in services and applications.

\bibliographystyle{iet-copy}
\bibliography{refs}

\end{document}